\documentclass[prc,showpacs,showkeys,superscriptaddress,nofootinbib,floatfix,onecolumn]{revtex4}
\usepackage{color}
\usepackage{amsfonts}
\usepackage{amsbsy}
\usepackage{mathrsfs}
\usepackage{graphicx}
\usepackage{subfigure}
\newcommand{\eeq}{\end{equation}}
\newcommand{\bea}{\begin{eqnarray}}
\newcommand{\eea}{\end{eqnarray}}

\def\lsim{\mathrel{\rlap{
\lower4pt\hbox{\hskip-3pt$\sim$}}
    \raise1pt\hbox{$<$}}}     %less than approx. symbol
\def\gsim{\mathrel{\rlap{
\lower4pt\hbox{\hskip-3pt$\sim$}}
    \raise1pt\hbox{$>$}}}     %greater than or approx. symbol

\begin{document}

%\DOIsuffix{theDOIsuffix}

%\Volume{XX}
%\Month{XX}
%\Year{XXXX}

%\pagespan{1}{}

%\Receiveddate{XXXX}
%\Reviseddate{XXXX}
%\Accepteddate{XXXX}
%\Dateposted{XXXX}

\title{Analytical contradictions \\ of the 'fixed - node' density matrix}
%}
\author{V.S.~Filinov}
\thanks{Corresponding author\quad E-mail:~\textsf{vladimir\_filinov@mail.ru}}
\affiliation{Joint Institute for High Temperatures, Russian Academy of
Sciences, Moscow, Russia}
%---------------------------
%\date{\today}% It is always \today, today,
             %  but any date may be explicitly specified

\begin{abstract} 
Over the last decades the 'fixed-node method' has been used for a 
numerical treatment of thermodynamic
properties of strongly correlated Fermi systems. In this work correctness of the 
'fixed -node method' for ideal Fermi systems has been analytically analyzed. 
It is shown that the 'fixed-node' prescription of calculation of the density 
matrix 
%Replacement of the initial condition for the Bloch equation by the zero boundary 
%condition in the 'fixed-node' calculation of the 
%density matrix 
leads to contradictions even for two ideal fermions.  
%Analogous contradictions 
%results from the virial decomposition of the many fermion 'fixed -- node' density %matrix.  
%matrix   obtained earlier in the 'fixed-node approach'.  
%Numerical results of the 'direct path integral Monte Carlo simulations' show that 
%the 'fixed -- node method' describes the thermodynamic properties of the strongly 
%coupled fermions rather well only at weak degeneracy. Difference in results %obtained 
%by these methods increases systematically with the growth of the degeneracy 
%at high density and low temperatures due to the wrong description of statistical %effects 
%by the 'fixed - node approach'.   
%Analogous conclusion  %\cite{cluster} 
%resulted from analytical cluster expansion of the many particle 
%density matrix of ideal Fermi system has been published by the author 
%eleven years ago. 
%So this method can not correctly describe the  ideal Fermi system. 
The main conclusion of this work is that the 'fixed-node method' can not reproduce 
the fermion density matrices and should be considered 
as uncontrolled empirical approach in treatment of thermodynamics of Fermi systems. \end{abstract}

\pacs{62.50.-p, 67.80.F-, 81.30.-t, 61.20.Ja}
\keywords{Fermi systems, fixed-node method, contradictions}

\maketitle

\section{Introduction}\label{s:intro} 

Over the last decades significant progress has been observed in theoretical studies of 
thermodynamic properties of strongly correlated fermions at non-zero temperatures,
which is mainly conditioned by the application of numerical simulations (see review 
\cite{Ceprl1}). The reason for this success is the possibility of an explicit representation of 
the %low-temperature
density matrix in the form 
%of a finite-dimensional approximation 
of the Wiener path integrals \cite{Feynm} and application of the Monte Carlo method for further 
calculations. The 
% without any preliminary physical approximations, which require some physical parameters to be small. 
main difficulty for path integral Monte Carlo (PIMC) studies 
of Fermi systems results from the requirement of antisymmetrization of the density matrix
% in the partition function 
 \cite{Feynm}. Then all thermodynamic quantities are 
presented as the sum of alternating sign terms related to even and odd permutations %of particles
and are equal to the small difference of two large numbers, 
which are the sums of positive and  negative terms. The numerical calculation in this case 
is severely hampered. This difficulty is known in the literature as the 'sign problem'. 
To overcome this issue some approaches have been developed, among which the 
'fixed-node method' \cite{Ceprl1,Ceprl2,Ceprl3,Militz} is widely known. 

To avoid sign problem at calculation of the path integral representation of the fermion density matrix the authors of \cite{Ceprl2,Ceprl3} suggested {\bf \em  'the path integral solution of the Bloch equation without minus signs'}. 
This means that they introduced restriction of integration  over 
paths by the domain, 
where additional 'trial antisymmetric density matrix'
 is not negative. The author of \cite{Ceprl2,Ceprl3} claimed that this 
 restriction of path integration gives the exact solution of the 
 Bloch equation in the form of path integrals with standard antisymmetric initial condition. 

The purpose of this work is to give an analytical proof that this restriction 
%of initial condition 
results in contradictions at explicit calculations of the density matrix even for two ideal fermions. 
The analogous contradictions have been analytically obtained twelve years ago in \cite{cluster} 
from  virial decomposition of the many fermion 'fixed-node' density matrix'. 
However paper \cite{cluster} is very difficult for understanding as used the Rueele algebraic 
approach \cite{Ruelle} and was missed by the scientific community. This is the reason to 
discuss correctness of the 'fixed - node method' once more using more simple mathematical technique. 
The main result of this work and paper \cite{cluster} is that 
the 'fixed--node method' can not reproduce even the well known 
ideal fermion density matrix and should 
be considered as an uncontrolled empirical approach in treatment 
of thermodynamics of fermions.  
%----------------------------

\section{Density matrix by the 'fixed - node method'} \label{fixed} 

Thermodynamic values of the fermion quantum system at non-zero temperature are defined by 
the appropriate derivatives of the logarithm of the partition function 
$Q_N = Tr \{ \rho \}$. Here $\rho=\exp(-\beta \hat{H})$ is the density matrix of 
a quantum system of particles with the Hamiltonian $\hat{H}=\hat{K}+\hat{U}$ equal to the sum of kinetic 
$\hat{K}$ and potential energy $\hat{U}$ operators, while $\beta = 1/k_B T$ .  
For our purposes it is enough to consider 
one dimensional (1D) system of two ideal fermions.  
So $\hat{U}\equiv0$, while the kinetic energy 
operator is the sum of two kinetic energy operators related to each particle  
 $\hat{K}=\hat{K_1}+\hat{K_2}$. 
Density matrix is the solution of the operator Bloch equation 
\begin{eqnarray}
\label{Bloch} 
\frac{\partial \hat{\rho}}{\partial \beta} = -\hat{K} \hat{\rho}
\end{eqnarray} 
with %$\hat{H}=\hat{K}$ and 
the initial condition $\hat{\rho} |_{\beta=0}=\hat{1}$. 

%According to \cite{Ceprl2,Ceprl3} 
This operator equation in coordinate representation for fermions can be written in the form 
\cite{Ceprl2,Ceprl3}   
\begin{eqnarray}
\label{Bloch1} 
\frac{\partial  \rho_{F}( R, R_0; \beta )}{\partial \beta} = -\hat{K}(R) \rho_{F}( R, R_0; \beta )
\end{eqnarray} 
with the initial condition 
\begin{eqnarray}
\label{InBl}
\rho_{F}( R, R_0; 0)=\frac{1}{N!}\sum_{P} (-1)^{\kappa P} \delta (R-PR_0) 
\end{eqnarray} 
where $R$ is the set of coordinates of all particles. 
One of the possible coordinate representation of the fermion density matrix looks like 
%can be written in the form  
\begin{eqnarray}
\label{Defin}
\rho_{F}( R, R_0; \beta ) = \frac{1}{N!}\sum_{\alpha} \exp(-\beta E_{\alpha}) 
\phi^{*}_{\alpha}(R) \phi_{\alpha}(R_0)
\end{eqnarray} 
where the sum is over the complete set of antisymmetric eigenfunctions 
$\phi_{\alpha}(R)$ of $\hat{H}$. 
%The sum in Eq.~(\ref{Defin}) is over all eigenfunctions irrespective of symmetry. 

Another {\bf \em exact} popular coordinate representation of operator $\hat{\rho}$ follows  
from the operator identity  
%\begin{eqnarray}
$%\label{Texp}
e^{-\beta \hat{K}}\equiv e^{-\Delta \beta {\hat K}} \cdots
e^{-\Delta \beta {\hat K}}\dots  e^{-\Delta \beta {\hat K}} 
$%\end{eqnarray}
{\bf \em for any (even of oder unity) integer fixed $M$}. Here the r.h.s. contains $M$ identical 
factors with $\Delta \beta = \beta/M$. So one can {\bf \em exactly} present the ideal density 
matrix in the form of finite-- difference expression of the path integral 
\begin{eqnarray}
\label{path}
\rho_{F}( R_M, R_0; \beta ) = \frac{1}{N!}\sum_{P} (-1)^{\kappa_P} \int dR_1 \cdots dR_{M-1}
\left(\prod_{k=1}^{M-1} \rho( R_{k-1}, R_k; \Delta \beta )\right)\rho( R_{M-1}, PR_M; \Delta \beta )
\end{eqnarray} 
where $N$ is the number of fermions. For $N=2$   arguments  
are two dimensional (2D) vectors composed by the coordinates of the first and second particle 
on $1D$ axes $X^{(1)}$ and $X^{(2)}$  and $\rho( R_{k-1}, R_k; \Delta \beta )$ 
are distinguishable particle density matrices. 
Antisymmetry is put in by the antisymmetrization.  The sum runs 
over all permutations with
parity $\kappa_P$ acting on indexes of particles. The density 
matrix is function of the space 2D 
arguments $R_M, R_0$ on ($X^{(1)}, X^{(2)}$) plane and inverse temperature 
(image time) $\beta$. Below we are going to discuss the boundary 
conditions of the certain domain on the ($X^{(1)}, X^{(2)}$) plane 
and mentioned before the initial conditions 
on $\beta$ of the parabolic equation (\ref{Bloch1}). 
 
\begin{figure}[htb]%\label{fig:cor}
\vspace{0cm} \hspace{0.0cm}
\includegraphics[width=6.5cm,clip=true]{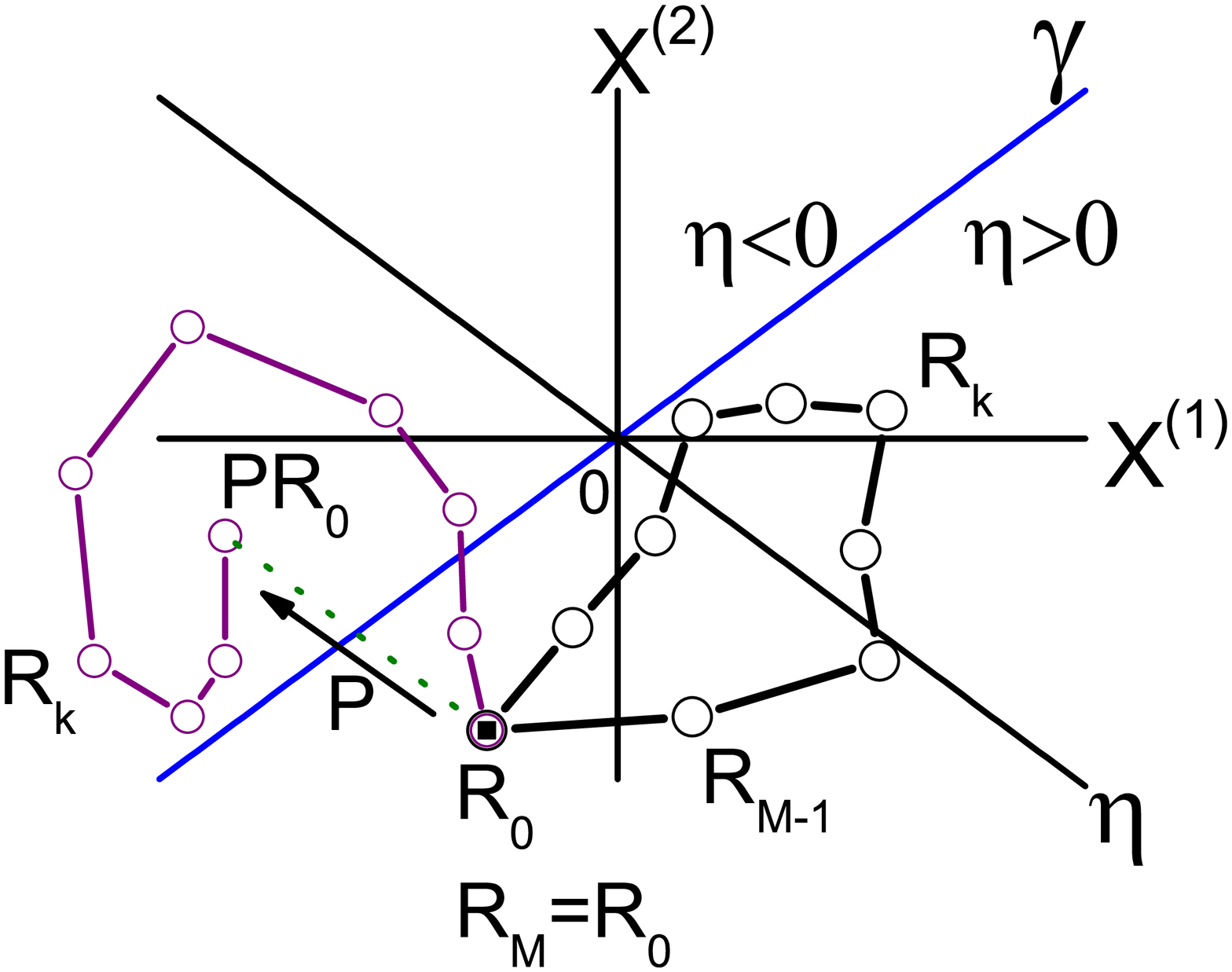}
\caption{(Color online) 
Configurational space of two free fermions. 
Plotted are the paths related to the two possible permutations: 
identical (below line $\gamma$) and non-identical (crossing line $\gamma$) permutations. 
}
\label{fig:GPath}
\end {figure}

Mathematical meaning of expression (\ref{path}) for density matrix  of two particles 
is illustrated by Fig. \ref{fig:GPath}, where vectors $R_k$ are presented by circles 
(called often as 'beads'), while 
density matrices $\rho( R_{k-1}, R_k; \Delta \beta )$ are denoted by segments of lines. 
Sometimes instead of coordinate $\left\{X^{(1)}_k,X^{(2)}_k\right\}$ it is convenient  
to use coordinates $\left\{\gamma_k,\eta_k\right\}$ defined by expressions 
$\gamma_k=0.5\left( X^{(1)}_k+X^{(2)}_k\right)$ and $\eta_k=\left( X^{(1)}_k-X^{(2)}_k\right)$, 
so $R_k=\left\{X^{(1)}_k,X^{(2)}_k\right\}=\left\{\gamma_k,\eta_k\right\}$.  
Modulus of Jacobian related to the change of variables of integration in Eq.~(\ref{path}) 
from the system of coordinates $\left\{X^{(1)}_k,X^{(2)}_k\right\}$ 
to the system $\left\{\gamma_k,\eta_k\right\}$ is equal to unity. % (see Fig. \ref{fig:GPath}).  
Action of perturbation $P$ is illustrated in Fig.~\ref{fig:GPath} by the arrow with letter 
$P$ (see Fig.~\ref{fig:GPath}). For two fermions the sum over permutations is reduced to 
the sum of contributions of identical and non identical permutations with opposite signs.   
%, $R_M=PR_0$
%Module of Jackobian related to the change of variables from $\left\{X^{(1)}_k,X^{(2)}_k\right\}$ 
%to $\left\{\gamma_k,\eta_k\right\}$ for integration in Eq.~(\ref{path}) is equal to unity. 
The density matrix has the following general properties: 
\begin{eqnarray}
\label{genpr}
\rho_{F}( R, R_0; \beta)=\rho_{F}(R_0, R ; \beta)=(-1)^{\kappa P}\rho_{F}(PR, R_0; \beta) 
\end{eqnarray} 

For further comparisons with the 'fixed - node' density matrix let us remind the solution 
of the Bloch equation (\ref{Bloch1}) with 
the initial condition  (\ref{InBl}). Density matrices in (\ref{path}) are well known 
for ideal particles and can be written in the form  \cite{Feynm} : 
\begin{eqnarray}
\label{twof}
\rho( R_{k-1}, R_k; \Delta \beta ) = 
\frac{\exp\left(-\frac{\pi|R_k-R_{k-1}|^2}{\tilde{\lambda}^2}\right)}{\tilde{\lambda}^2}=
\frac{\exp\left(-\frac{2\pi|\gamma_k-\gamma_{k-1}|^2}{\tilde{\lambda}^2}\right)
\exp\left(-\frac{\pi|\eta_k-\eta_{k-1}|^2}{2\tilde{\lambda}^2}\right)}{\tilde{\lambda}^2}
\end{eqnarray}  
where $\tilde{\lambda}^2=2\pi\hbar^2\Delta \beta/m$ is the thermal wave length related to 
$\Delta \beta$. 
The last factor in (\ref{path}) for permutation $P$ has the form 
\begin{eqnarray}
\label{twof2}
\rho( R_{M-1}, PR_M; \Delta \beta ) = 
\frac{\exp\left(-\frac{\pi|PR_M-R_{M-1}|^2}{\tilde{\lambda}^2}\right)}{\tilde{\lambda}^2}=
\frac{\exp\left(-\frac{2\pi|P\gamma_M-\gamma_{M-1}|^2}{\tilde{\lambda}^2}\right)
\exp\left(-\frac{\pi|P\eta_M-\eta_{M-1}|^2}{2\tilde{\lambda}^2}\right)}{\tilde{\lambda}^2}.
\end{eqnarray}  
So the exact well known expression for two particle  antisymmetrized  density matrix looks like \cite{Feynm} 
\begin{eqnarray}
\label{twof3}
&&\rho_{F}( R_M, R_0; \beta )=\frac{1}{2}
\int dR_1 \cdots dR_{M-1}
\left(\prod_{k=1}^{M-1} \frac{\exp\left(-\frac{\pi|R_k-R_{k-1}|^2}{\tilde{\lambda}^2}\right)}
{\tilde{\lambda}^2}\right)
\frac{\exp\left(-\frac{2\pi|\gamma_M-\gamma_{M-1}|^2}{\tilde{\lambda}^2}\right)}{\tilde{\lambda}}
\nonumber\\&& \times 
%\nonumber\\&\times& 
\left\{
\frac{\exp\left(-\frac{\pi|\eta_M-\eta_{M-1}|^2}{2\tilde{\lambda}^2}\right)-
\exp\left(-\frac{\pi|P\eta_M-\eta_{M-1}|^2}{2\tilde{\lambda}^2}\right)}{\tilde{\lambda}}
%\rho( R_{M-1}, R_M; \Delta \beta)-\rho(R_{M-1}, PR_M; \Delta \beta)
\right\}
\nonumber\\&\equiv& \frac{1}{2} 
\frac{\exp\left(-\frac{2\pi|\gamma_M-\gamma_{0}|^2}{\lambda^2}\right)}{\lambda}
%\nonumber\\&& 
%\nonumber\\&\times& 
\left\{
\frac{\exp\left(-\frac{\pi|\eta_M-\eta_{0}|^2}{2\lambda^2}\right)-
\exp\left(-\frac{\pi|P\eta_M-\eta_{0}|^2}{2\lambda^2}\right)}{\lambda}
%\rho( R_{M-1}, R_M; \Delta \beta)-\rho(R_{M-1}, PR_M; \Delta \beta)
\right\}
\end{eqnarray} 
%with $P\gamma_M=\gamma_M$ and $P\eta_M=-\eta_M$. If $R_M=R_0$ then 
%$P\gamma_M=\gamma_0$ and $P\eta_M=-\eta_0$.  
with $\lambda^2=2\pi\hbar^2\beta/m$, $P\gamma_M=\gamma_M$ and $P\eta_M=-\eta_M$. 
If $\left\{\gamma_M,\eta_M\right\}=\left\{\gamma_0,\eta_0\right\}$ then 
$P\gamma_M=\gamma_0$ and $P\eta_M=-\eta_0$ (see Fig.~\ref{fig:GPath}).  

To avoid 'sign problem' at calculation of the path integral representation 
of the fermion density matrix the authors of \cite{Ceprl2,Ceprl3} suggested 
the 'fixed - node' {\bf \em  'the path integral solution of the 
Bloch equation without minus signs'}. 
The authors of \cite{Ceprl2,Ceprl3} denotes the second argument 
$R_0$ of the density matrix 
as the {\bf \em 'reference point'} and {\bf \em 'the set of points $R_t$ for which 
there exists a continuous 'space--imaginary time' path with $\rho_{F}( R, R_0; t' )>0$ 
for $0\leq t' \leq \beta $ the reach of $R_0$ or $\Gamma(R_0, \beta)$'}. 
For two ideal fermions the reach can be analytically obtained \cite{Ceprl2,Ceprl3}. 
The reach (the half plane $\eta>0$) for $\eta_0>0$ is shown in Fig.~\ref{fig:GPath}.  
The reach does not depend on temperature. 

%To avoid the sign problem at calculation of the path integral representation of %the fermion density matrix the authors of \cite{Ceprl2,Ceprl3} suggested {\bf \em  %'the path integral solution of the Bloch equation without minus signs'}.  
According to the papers \cite{Ceprl2,Ceprl3} 
{\bf \em 'It is a simple matter (see Appendix C) to show that the problematic 
INITIAL CONDITION, Eq.~(\ref{InBl}), can be replaced by a zero 
boundary conditions on the SURFACE of the reach.  It follows 
because the fermion density matrix is a unique solution to the Bloch 
equation (\ref{Bloch1}) with the zero BOUNDARY CONDITION.'} 
However the Bloch equation  (\ref{Bloch1}) with the zero boundary conditions is 
linear equation and, besides the trivial solution identically equal to zero, 
has an infinite number of solutions distinguishing at least by constant factors, 
 while the Bloch equation  (\ref{Bloch1}) with  initial condition 
 (\ref{InBl}) has really a unique solution.  

Let us consider 'fixed -- node' approach to calculation of the 
density matrices in the 
$\gamma - \eta$ plane (see Fig. \ref{fig:GPath}). 
The author of \cite{Ceprl2,Ceprl3} claimed that to obtain the 'fixed -- node' 
density matrix  {\bf \em 
'one simply restricts the paths in 
Eq.~(\ref{path}) to lie in the  $\Gamma(R_0, \beta)$'}. 
%{\bf \em 'it follows because the fermion density matrix is 
%a unique solution to the Bloch equation (\ref{Bloch1}) 
%with the zero boundary conditions (see Appendix C)'}
This means that 
restriction of integration  over 
$R_1,\cdots,R_{M-1}$ in (\ref{path}) {\bf \em by the reach} 
has to give the exact solution of 
Eq.~(\ref{Bloch1}) with initial condition (\ref{InBl}).   
%The background of papers \cite{Ceprl2,Ceprl3} is in reformulation of antisymmetry in terms of a boundary 
%condition.   
The fallaciousness of this statement  for $M=2$ as well as for arbitrary integer $M$ 
can be easily proved by explicit integration over 
variables $R_1,\cdots,R_{M-1}$ in the reach for 1D two ideal fermions. The boundary surface of the 2D reach for two ideal fermions is exactly known 
and according to papers 
\cite{Ceprl2,Ceprl3} is the line $\eta=0$ in Fig.~\ref{fig:GPath}.  
So according to the 'fixed-node' prescription  
the density matrix in the whole 
configurational space (in both half planes 
($\eta>0$ and $\eta<0$)  ($\eta_M, \eta_0\in \{X^{(1)}, X^{(2)}\}$)) 
% of Eq.~(\ref{Bloch1}) 
%with zero boundary conditions ($\eta_M=0$) 
looks like 
%$\left\{\gamma_M,\eta_M\right\}=\left\{\gamma_0,\eta_0\right\}$
\begin{eqnarray}
\label{twoff}
&&\rho_{F}( R_M, R_0; \beta )=\frac{C_1}{2}
\int dR_1 \cdots dR_{M-1}
\left(\prod_{k=1}^{M-1} \frac{\exp\left(-\frac{\pi|R_k-R_{k-1}|^2}{\tilde{\lambda}^2}\right)}
{\tilde{\lambda}^2}\right)
\frac{\exp\left(-\frac{2\pi|\gamma_M-\gamma_{M-1}|^2}{\tilde{\lambda}^2}\right)}{\tilde{\lambda}}
\nonumber\\&& \times
%\nonumber\\&\times& 
\left\{
\frac{\exp\left(-\frac{\pi|\eta_M-\eta_{M-1}|^2}{2\tilde{\lambda}^2}\right)-
\exp\left(-\frac{\pi|P\eta_M-\eta_{M-1}|^2}{2\tilde{\lambda}^2}\right)}{\tilde{\lambda}}
%\rho( R_{M-1}, R_M; \Delta \beta)-\rho(R_{M-1}, PR_M; \Delta \beta)
\right\}
\theta(\eta_0)\theta(\eta_1)\cdots\theta(\eta_{M})
\nonumber\\&& +
\frac{C_2}{2}
\int dR_1 \cdots dR_{M-1}
\left(\prod_{k=1}^{M-1} \frac{\exp\left(-\frac{\pi|R_k-R_{k-1}|^2}{\tilde{\lambda}^2}\right)}
{\tilde{\lambda}^2}\right)
\frac{\exp\left(-\frac{2\pi|\gamma_M-\gamma_{M-1}|^2}{\tilde{\lambda}^2}\right)}{\tilde{\lambda}}
\nonumber\\&& \times
%\nonumber\\&\times& 
\left\{
\frac{\exp\left(-\frac{\pi|\eta_M-\eta_{M-1}|^2}{2\tilde{\lambda}^2}\right)-
\exp\left(-\frac{\pi|P\eta_M-\eta_{M-1}|^2}{2\tilde{\lambda}^2}\right)}{\tilde{\lambda}}
%\rho( R_{M-1}, R_M; \Delta \beta)-\rho(R_{M-1}, PR_M; \Delta \beta)
\right\}
\theta(-\eta_0)\theta(-\eta_1)\cdots\theta(-\eta_{M}),
\end{eqnarray} 
where $\theta(\eta)$ is the theta function 
equal to zero for $\eta < 0$ and equal to unity in 
the opposite case $\eta \geq 0$ . The theta functions restrict 
to the reach the domains of integration. 
Now assume that all 'basic statements' of papers \cite{Ceprl2,Ceprl3} 
are {\bf \em correct}, then using 
the mentioned above general properties of the density matrix we have to admit that integration over 
$R_2\cdots,R_{M-1}$ in the reach gives the exact solution of Eq.~(\ref{Bloch1}) with initial 
condition (\ref{InBl}). So the density matrix in the 'fixed--node method' (\ref{twoff}) 
can be transformed to the following integral over the last variable $R_1$: 
\begin{eqnarray}
\label{twofff}
&&\rho_{F}( R_M, R_0; \beta )=\frac{1}{2} 
\int dR_1 %\cdots dR_{M-1}
%\left(\prod_{k=1}^{M-1} 
\frac{\exp\left(-\frac{\pi|R_1-R_{0}|^2}{\tilde{\lambda}^2}\right)}
{\tilde{\lambda}^2}
%\right)
\frac{\exp\left(-\frac{2\pi|\gamma_M-\gamma_{1}|^2}{(M-1)\tilde{\lambda}^2}\right)}{\sqrt{(M-1)}\tilde{\lambda}}
%\nonumber
\\&&\times 
%\nonumber\\&\times& 
[C_1%\left\{ 
\frac{\exp\left(-\frac{\pi|\eta_M-\eta_{1}|^2}{2(M-1)\tilde{\lambda}^2}\right)-
\exp\left(-\frac{\pi|P\eta_M-\eta_{1}|^2}{2(M-1)\tilde{\lambda}^2}\right)}{\sqrt{(M-1)}\tilde{\lambda}}
%\rho( R_{M-1}, R_M; \Delta \beta)-\rho(R_{M-1}, PR_M; \Delta \beta)
%\right\}
\theta(\eta_0)\theta(\eta_1)\theta(\eta_M)
\nonumber\\&&+
%\frac{C_1}{2} 
%\int dR_1 %\cdots dR_{M-1}
%%\left(\prod_{k=1}^{M-1} 
%\frac{\exp\left(-\frac{\pi|R_1-R_{0}|^2}{\tilde{\lambda}^2}\right)}
%{\tilde{\lambda}^2}
%\right)
%\frac{\exp\left(-\frac{2\pi|\gamma_M-\gamma_{1}|^2}{(M-1)\tilde{\lambda}^2}\right)}{\sqrt{(M-1)}\tilde{\lambda}}
%\nonumber\\&& \times 
%\nonumber\\&\times& 
%\left\{
C_2\frac{\exp\left(-\frac{\pi|\eta_M-\eta_{1}|^2}{2(M-1)\tilde{\lambda}^2}\right)-
\exp\left(-\frac{\pi|P\eta_M-\eta_{1}|^2}{2(M-1)\tilde{\lambda}^2}\right)}{\sqrt{(M-1)}\tilde{\lambda}}
%\rho( R_{M-1}, R_M; \Delta \beta)-\rho(R_{M-1}, PR_M; \Delta \beta)
\theta(-\eta_0)\theta(-\eta_1)\theta(-\eta_M)] %\right\} 
%\nonumber\\&& = 
\nonumber\\&&=  
\frac{1}{2} \frac{\exp\left(-\frac{2\pi|\gamma_M-\gamma_0|^2}{\lambda^2}\right)}
{2\lambda}
%\nonumber\\&& \times
%\left\{
[C_1\theta(\eta_0)\theta(\eta_M)
\nonumber\\&&\times 
\frac{\exp\left(-\frac{\pi|\eta_M-\eta_{0}|^2}{2\lambda^2}\right)
\xi((M-1)\eta_{0}+\eta_M)-
\exp\left(-\frac{\pi|P\eta_M-\eta_{0}|^2}{2\lambda^2}\right)
\xi((M-1)\eta_{0}+P\eta_M)
}{\lambda}
%\right\}
\nonumber\\&&+
%\left\{
C_2\theta(-\eta_0)\theta(-\eta_M)
\nonumber\\&&\times
\frac{\exp\left(-\frac{\pi|\eta_M-\eta_{0}|^2}{2\lambda^2}\right)
\xi(-(M-1)\eta_{0}+\eta_M)-
\exp\left(-\frac{\pi|P\eta_M-\eta_{0}|^2}{2\lambda^2}\right)
\xi(-(M-1)\eta_{0}+P\eta_M)
}{\lambda}
]%\right\}
\nonumber
\end{eqnarray} 
where $\xi\left(z\right)=erfc\left(-\frac{\sqrt{\pi}z}{\sqrt{2M(M-1)}\tilde{\lambda}}\right)$

Thus instead of the unique density matrix (\ref{twof3}) 
we have infinite number of the 
'fixed -- node' density matrices (\ref{twofff}) (due to the two 
arbitrary constants $C_1$ and $C_2$) taking the 'zero boundary conditions' 
on the surface of the reach ($\eta_M=0$) for any finite $M\geq 2$. 
More over this density matrix 
 depend on complementary error functions. 
% depending on any (even of order unity) $M$.  
%Note that the obtained expression for the 
%'fixed--node' density matrix is defined to arbitrary factor $C_1$. This expression 
%for 'fix-node' density matrix Eq.~
%This is the exact general 'fixed-node' solution for 
%any fixed $M\geq 2$. 
%and is very simple for $M=2$.  

For further detail analysis of the function  (\ref{twofff}) let us consider the limit of 
$\Delta \beta \rightarrow 0$. % ($M \rightarrow  \infty$). 
%, which looks much more easier. 
Using definition of complementary error functions one can transform (\ref{twofff}) to the form: 
\begin{eqnarray}
\label{threefff}
&&\rho_{F}( R_M, R_0; \beta )= 
\frac{\exp\left(-\frac{2\pi|\gamma_M-\gamma_0|^2}{\lambda^2}\right)}
{4\lambda}
\nonumber\\&& \times
%\left\{
[C_1\theta(\eta_0)\theta(\eta_M)
\frac{(\exp\left(-\frac{\pi|\eta_M-\eta_{0}|^2}{2\lambda^2}\right)-
\exp\left(-\frac{\pi|P\eta_M-\eta_{0}|^2}{2\lambda^2}\right))
(1+sign(\eta_0))
}{\lambda}
%\right\}
\nonumber\\&& + 
%\left\{
C_2\theta(-\eta_0)\theta(-\eta_M)
\frac{(\exp\left(-\frac{\pi|\eta_M-\eta_{0}|^2}{2\lambda^2}\right)-
\exp\left(-\frac{\pi|P\eta_M-\eta_{0}|^2}{2\lambda^2}\right))
(1-sign(\eta_0))
}{\lambda}]
\nonumber\\&& =
\frac{\exp\left(-\frac{2\pi|\gamma_M-\gamma_0|^2}{\lambda^2}\right)}
{4\lambda}
\frac{(\exp\left(-\frac{\pi|\eta_M-\eta_{0}|^2}{2\lambda^2}\right)-
\exp\left(-\frac{\pi|P\eta_M-\eta_{0}|^2}{2\lambda^2}\right))
%(1+sign(\eta_0))
}{\lambda}
%\nonumber\\&& \times
%\left\{
\nonumber\\&& \times 
[(C_1\theta(\eta_0)\theta(\eta_M)+C_2\theta(-\eta_0)\theta(-\eta_M))
\nonumber\\&& +
(C_1\theta(\eta_0)\theta(\eta_M)-C_2\theta(-\eta_0)\theta(-\eta_M))
sign(\eta_0)]
%\nonumber
\end{eqnarray} 
where 
\begin{eqnarray}
\label{th}
&&
lim_{\Delta \beta \rightarrow 0} erfc\left(-\sqrt{\pi}\frac{\pm(M-1)\eta_{0}+P\eta_M}
{\sqrt{2M(M-1)}\tilde{\lambda}}\right)
\nonumber\\&& = 
1-lim_{\Delta \beta \rightarrow 0} 
 erf\left(-\sqrt{\pi}\frac{\pm(M-1)\eta_{0}+P\eta_M}
{\sqrt{2M(M-1)}\tilde{\lambda}}\right)= 
1\pm sign(\eta_0)
\end{eqnarray} 
and $erf$ is the error function. Here $sign(\eta_{0})$ is equal to plus unity for $\eta_{0}>0$ 
and minus unity for $\eta_{0}<0$. 
In the limit $\Delta \beta \rightarrow 0$ 
the 'fixed--node' density matrix (\ref{threefff}) for two arbitrary constants 
($C_1$ and $C_2$) takes the 'zero boundary conditions' on the surface of the reach ($\eta_M=0$). 

To obtain the unique density matrix we need to specify 
constants $C_1$ and $C_2$. 
The 'fix-node' density matrix (\ref{threefff}) differs significantly 
from  exact expressions Eq.~(\ref{twof3}). 
For $C_1=1$ and $C_2=0$ 
we have density matrix, which 
coincide with exact density matrix if both $\eta_M$ and $\eta_0$ 
are positive but if  $\eta_M$ and $\eta_0$ have opposite sign 
the 'fixed-node' density matrix is identically equal to zero 
and differs from exact density matrix. 
Generally the 'fix-node' density matrix (\ref{threefff}) 
is identically equal to zero if $\eta_0$ and $\eta_M$ 
are lying in opposite half planes ($\eta>0$ and $\eta<0$) 
of the $\gamma - \eta$ plane, 
 while this is not the case for exact density matrix (\ref{twof3}). 
  The 'fix-node' density matrix (\ref{threefff})  contrary to 
the exact one is {\bf \em a non analytical function}. 
Let us remind that all these expressions have been obtained 
at assumption that all 'basic statements' of papers 
\cite{Ceprl2,Ceprl3} are {\bf \em correct}. 
%The 'fix-node' density matrix (\ref{threefff})  contrary to 
%the exact one is {\bf \em a non analytical function} with 
%the following 'initial conditions' at $\beta=0$: 
%\begin{eqnarray}
%\label{inifxnd}
%&&\rho_{F}( R_M, R_0; \beta=0 )= 
%%\nonumber\\&& =
%\frac{\delta\left(\gamma_M-\gamma_{0}\right)
%(\delta\left(\eta_M-\eta_{0}\right)
%-\delta\left(P\eta_M-\eta_{0}\right))}{4}
%\nonumber\\&& \times
%[(C_1\theta(\eta_0)\theta(\eta_M)+C_2\theta(-\eta_0)\theta(-\eta_M))
%\nonumber\\&& +
%(C_1\theta(\eta_0)\theta(\eta_M)-C_2\theta(-\eta_0)\theta(-\eta_M))
%sign(\eta_0)]
%%\nonumber\\&&
%%[(C_1+C_2) 
%%%(\delta\left(\eta_M-\eta_{0}\right)-
%%%\delta\left(P\eta_M-\eta_{0}\right))
%%%\nonumber\\&& 
%%%+ 
%%(C_1\theta(\eta_M)-C_2\theta(-\eta_M)) 
%%sign(\eta_0)]
%%%\right\}
%%%\nonumber
%\end{eqnarray} 
%Let us consider the {\bf \em 'initial conditions' } at $\beta=0$  
%of the 'fixed-node' density matrix. 
%Expression (\ref{inifxnd}) 
%do not coincide with standard initial conditions (\ref{InBl})
%and depend on the $C_1$ and $C_2$. Let us note that 
%it is impossible to chose constants $C_1$ and $C_2$ in the 
%'fix-node initial conditions' (\ref{inifxnd}) to repoduce the 
% standard initial conditions 
%(\ref{InBl}), so the 'fixed-node' density matrix {\bf \em can not be solution 
%of the Bloch equation (\ref{Bloch1})  with initial condition  (\ref{InBl})}.  
%The 'fix-node' density matrix for 
%negative $\eta_0$ is identically equal to zero. 
All these contradictions mean that the integration in the reach can not reproduce 
the exact solution of Eq.~(\ref{Bloch1}) with initial 
condition (\ref{InBl}) in spite of the 'basic statement' of papers  \cite{Ceprl2,Ceprl3}. 

An alternative approach for studies Fermi systems without replacement of initial conditions 
by zero boundary conditions for the Bloch equation is known in literature as the 
▒direct path integral Monte Carlo simulation▓ (DPIMC) \cite{cluster,Egger,Imada,FBF,FFBK}.  
%[10, 16√20]. 
In this approach the sum over all permutations is represented identically as a determinant, which can 
be exactly calculated by the direct methods of linear algebra. The accuracy of this approach 
depends only on the errors of the finite-dimensional approximations of the path integrals 
and can be improved systematically. 
Comparison with results of the DPIMC simulation show that the ▒fixed -- node method▓ describes the 
thermodynamic properties of the {\bf strongly coupling} fermions rather well at 
weak degeneracy, when the main contribution to the partition function comes from the 
identical permutation \cite{cluster,FBF,FFBK}. The difference in obtained results increases 
systematically with the growth of the degeneracy at high density and low temperatures 
\cite{cluster,FBF,FFBK}. 
The reason of this difference is in 
restriction by the 'reach' of integration over 'beads' in 
the 'fixed - node' path integrals, which leads 
to wrong expression for density matrix even for two ideal fermions. 
This restriction results in uncontrolled errors in calculations of 
thermodynamic quantities  due to the wrong description of 
statistical effects in the system of  
degenerate interacting and non interacting fermions . 

\section{Conclusion}

Let us sum up analytical contradictions following from the basic 
'fixed -- node' prescription of calculation of the fermion density matrix 
({\bf \em 'one simply restricts the paths in Eq.~(\ref{path}) to lie in the  $\Gamma(R_0, \beta)$'} \cite{Ceprl2,Ceprl3}): 

1) instead of the unique density matrix (\ref{twof3}) the 
'fixed -- node' approach  gives infinite number 
of the density matrices (\ref{twofff})  
taking the 'zero boundary conditions' on the surface of 
the reach ($\eta_M=0$) for any {\bf \em finite} number of 'beads' $M\geq 2$  
(due to the two arbitrary constants $C_1$ and $C_2$); 

2)to obtain any unique solution (to define $C_1$ and $C_2$) 
the 'fixed-node' path integral restriction  
have to be supplemented with any additional condition, which does not discussed 
in \cite{Ceprl2,Ceprl3} (except the zero initial condition leading 
to the the trivial solution of the parabolic differential equation 
identically equal to zero (Appendix C formula (C2))); 

3) for finite $M$ the 'fixed--node' density matrix (\ref{threefff}) taking the 'zero boundary conditions' on the surface of the reach ($\eta_M=0$) differs from 
exact  density matrix for {\bf \em any choice} of the constants $C_1$ and $C_2$;  

4)the 'fixe-node'  density matrices  depend on $M$ and complementary error functions, which is not the case for exact one; 

5)in the limit $M \rightarrow \infty$ ($\Delta \beta \rightarrow 0$) 
the 'fixed--node' density matrix (\ref{threefff}) taking the 'zero boundary
 conditions' on the surface of the reach ($\eta_M=0$) is not unique and 
 differs from exact  density matrix for {\bf \em any choice} of the 
 constants $C_1$ and $C_2$;  

6)the 'fix-node' density matrices (\ref{twofff}) and (\ref{threefff}) contrary to 
exact expressions Eq.~(\ref{twof3})  is {\bf \em a non analytical function 
and therefore can not be solution 
of the Bloch equation (\ref{Bloch1})  with initial condition  (\ref{InBl})}. ; 
%
%4) the {\bf \em 'initial conditions' } at $\beta=0$  of the 
%'fixed-node' density matrix do not coincide with standard 
%initial conditions (\ref{InBl}) and depend on the $C_1$ and $C_2$; 
% 
%5) it is impossible to chose constants $C_1$ and $C_2$ in the 
%'fix-node initial conditions' to get standard initial conditions 
%(\ref{InBl}), so the 'fixed-node' density matrix {\bf \em can not be solution 
%of the Bloch equation (\ref{Bloch1})  with initial condition  (\ref{InBl})}.  

So the 'fixed - node method' can not reproduce correctly 
even the two fermions density matrix. 
Analogous conclusion for the many particle density matrix of ideal Fermi system  %\cite{cluster} 
have been analytically obtained in  \cite{cluster} from   virial decomposition. 
So the 'fixed - node method' can not correctly describe the  degenerate Fermi systems. 
The main result of this simple work and paper \cite{cluster} is that the 'fixed - node method' should 
be considered as {\bf \em uncontrolled} empirical approach in treatment of thermodynamics of 
degenerate non interacting fermions.  Numerical simulations of thermodynamic quantities 
for interacting fermions by the direct path integral Monte Carlo method results in analogous conclusions. 
%----------------------------

%----------------------------
\subsection*{Acknowledgements}

Author acknowledge stimulating discussions with academician V.E.~Fortov 
and Profs. P.R.~Levshov and S.Ya.~Bronin.
%
%----------------------------

\bibliographystyle {apsrev}

\begin{thebibliography}{90}
%%
%%1
%3
\bibitem{Ceprl1}
%The properties of hydrogen and helium under extreme conditions. 
J~M~McMahon, M~A~Morales, C~Pierleoni, D~Ceperley, 
Rev. Mod. Phys. {\bf 84}, 1607 (2012) 
%4
\bibitem{Feynm}
R~P~Feynman and A~R~Hibbs,  {\it Quantum
Mechanics and Path Integrals} (New York: McGraw-Hill)(1965).
% 
\bibitem{Ceprl2}
D~Ceperley J. Stat. Phys. {\bf 63}, 1237 (1991) 
%
\bibitem{Ceprl3}
D~Ceperley, Phys. Rev. Let. {\bf 69}, 331 (1992)
%
\bibitem{Militz}
B~Militzer and R~Pollock Phys. Rev. E,  {\bf 61}, 3470 (2000)
%
\bibitem{cluster}
V~Filinov J. Phys. A: Math. Gen.  {\bf 34}, 1665 (2001) 

\bibitem{Ruelle}
D~Ruelle  {\it Statistical Mechanics. Rigorous results} 
(New York: Benjamin-Cummings) (1969)

\bibitem{Egger} 
R~Egger, W~Hausler, C~H~Mak and H~Grabert Phys. Rev. Lett. {\bf 82}, 
3320 and references therein (1999)

\bibitem{Imada} 
R~Imada  J. Phys. Soc. Japan. {\bf 53} 2861 ( 1984)

\bibitem{FBF} 
V~S~Filinov, M~Bonitz and V~E~Fortov,  JETP Lett. {\bf 72}, 245 (2000)

\bibitem{FFBK}
V~S~Filinov, V~E~Fortov, M~BonitzM and D~Kremp, 
Phys. Lett. A {\bf274} 228 (2000)

\end{thebibliography}

\end{document}